\documentclass[aps,prb,twocolumn,superscriptaddress,longbibliography]{revtex4-1}
\usepackage{epsfig,amsopn}
\usepackage{graphicx}
\usepackage{color}
\usepackage{amsmath,amssymb}
\usepackage{enumerate}
\newcommand\bea{\begin{eqnarray}}
\newcommand\eea{\end{eqnarray}}
\newcommand\beq{\begin{equation}}
\usepackage{gensymb}
\usepackage{verbatim}
\newcommand\eeq{\end{equation}}

\def\nn{\nonumber}
\def\f{\frac}

\def\si{\sigma}
\def\Do{\partial}
\def\De{\Delta}

\def\ua{\uparrow}
\def\da{\downarrow}

\def\th{\theta}

\begin{document}
\title{Crossed Andreev reflection in altermagnets} 
\author{Sachchidanand Das}
\affiliation{School of Physics, University of Hyderabad, Prof. C. R. Rao Road, Gachibowli, Hyderabad-500046, India}

 \author{ Abhiram Soori}
 \email{abhirams@uohyd.ac.in}
 \affiliation{School of Physics, University of Hyderabad, Prof. C. R. Rao Road, Gachibowli, Hyderabad-500046, India}
\begin{abstract}
Crossed Andreev reflection (CAR) is a  scattering phenomenon occurring in a superconductor (SC) connected to two metallic leads, where an incident electron on one side of the SC emerges on the opposite side as a hole. Despite its significance, CAR detection is often impeded by the prevalent electron tunneling (ET), wherein the incident electron  exits on the opposing side as an electron. One approach to augment CAR over ET involves employing two antiparallel ferromagnets across the SC. However, this method is constrained by the low polarization in ferromagnets and necessitates the application of a magnetic field. Altermagnets (AMs) present a promising avenue for detecting and enhancing CAR due to their distinct Fermi surfaces for the two spins. Here, we propose a  configuration utilizing two AMs rotated by $90^{\circ}$ with respect to each other on either side of an SC to enhance CAR. We calculate local and nonlocal conductivities across the AM-SC-AM junction using the Landauer-B\"uttiker scattering approach. Our findings reveal that in the strong phase of AMs, CAR overwhelmingly dominates nonlocal transport. In the weak phase, CAR can exhibit significant enhancement for larger values of the altermagnetic parameter compared to the scenario where AMs are in the normal metallic phase. As a function of the length of the SC, the conductivities exhibit oscillations reminiscent of Fabry-P\'erot interference. 
\end{abstract}
\maketitle
\section{Introduction}

Electrons are characterized by charge and spin. While the charge of the electrons is a characteristic used in electronics, their spin is the characteristic which has opened up a new field of study known as spintronics~\cite{wolf01,chappert07,baltz18,fukami20,hoffmann22,sinova04,bijay22}. Ferromagnets and  antiferromagnets have played a central role in spintronics so far\cite{baltz18}. In ferromagnets, a majority of  spins are aligned in one  direction, resulting in a net spin polarization. On the other hand, in antiferromagnets neighboring spins point in opposite direction, making the net spin polarization zero. Recently, a new class of magnetic materials known as AMs have generated interest among theorists and experimentalists~\cite{smejkal22a,smejkal22b,smejkal22c,fern24,sun23ar,zhou24,yan24,reich24}. In AMs, the dispersions of the two spin sectors are separated in momentum space while maintaining a zero net spin polarization. In contrast to spin-orbit coupled systems, time reversal symmetry is broken in AMs. A consequence of such feature is that even though the net spin polarization is zero in both: a metal and an AM, a junction between the two carries a net spin current on application of a voltage bias~\cite{das2023}. Even before AMs were introduced, the concept of spin split bandstructures as in AMs was  explored in literature~\cite{hayami19,hayami20}. It has been predicted that some antiferromagnets can be turned into AMs by  application of electric field~\cite{mazin23induced}. 

Even in ferromagnetic metals (FMs), the dispersions for the two spins are separated, but with a net spin polarization. This property enables the use of FMs in detecting crossed Andreev reflection (CAR)~\cite{beckmann04} - a phenomenon wherein an electron incident onto a SC from a FM gets transmitted into another FM as a hole. Enhancement of CAR is important from the point of view of developing quantum devices that use non-locally entangled electrons. Cooper pairs in singlet SC are entangled. Cooper pair splitting (CPS) is a process which separates the two electrons of a Cooper pair into two metals connected to the SC, maintaining their entanglement~\cite{recher2001,Das2012,schin12}. CPS happens when current is driven from SC into two connected metals, whereas CAR happens when the current is driven from one of the two metals into the SC. Hence, CPS is the inverse process of CAR and enhancing CAR in a setup can result in an enhanced CPS in the same setup when the current is driven from SC to the two metals. 

Hindrance to experimental observation of CAR is rooted in a competing process known as electron tunneling (ET) in which the electron transmits across the SC from one metal onto the other metal as an electron. The currents carried by ET and CAR are opposite in sign and in most cases, the current carried by ET overpowers the current carried by CAR and masks the signature of CAR~\cite{cht}. Several proposals have been put forward to circumvent this limitation and enhance CAR over ET~\cite{melin2004sign,sadovskyy2015,soori17,nehra19,soori22car,Zhang19,Jakobsen21,Zhao}, of which two methods have been implemented experimentally~\cite{beckmann04,russo2005experimental} including the one where  two antiparallel  FMs are used. The use of FMs typically requires application of magnetic fields in experiments~\cite{beckmann04,Valenzuela2006}. Antiparallel alignment of ferromagnetic electrodes requires an external field, carefully stabilized in a particular field region~\cite{moodera1995}. In devices that comprise CAR based components, applying magnetic field will affect the functioning of other components.  Effect of ferromagnetic components along with external magnetic field can cause significant effects due to spurious fringing fields\cite{moodera23}, which need to be eliminated - hence absence of external magnetic field/ FMs is an advantage. While a magnetic field of the order of $20$mT is needed to get the FMs antiparallel~\cite{beckmann04}, a magnetic field of the order of $0.1$mT is sufficient to give rise to an effect such as superconducting diode effect~\cite{moodera23}. 
 AMs provide an edge over FMs, because they require no external field for such alignment~\cite{smejkal22c}. 

AMs have been predicted to affect the superconducting transition temperature when coupled with SCs~\cite{giil23}. In this work~\cite{giil23}, SC  sandwiched between two AMs in two possible configurations is studied: the two AMs are parallel, and the two AMs are rotated by $90^{\circ}$ with respect to one another.  The superconducting transition temperature is found to be lower in the latter configuration compared to that in the former. The authors claim that this is due to enhanced CPS (inverse CAR) in the latter configuration, which favors the Cooper pairs to be broken into electrons. The idea that CAR is enhanced when the two AMs rotated with respect to one another are  connected to an SC is worth an in-detail investigation. Also, Josephson effect that depends on the crystallographic orientation in SC-AM-SC junctions have been studied recently~\cite{Ouassou23,Beenakker23,Cheng24}.

In this work, we study transport across an AM-SC-AM junction. We find that CAR is enhanced for certain crystallographic orientations of  AMs. For AMs in the strong phase, ET can be completely suppressed, making way for CAR to solely dominate the nonlocal transport.

\begin{figure}[htb]
 \includegraphics[width=8.0cm]{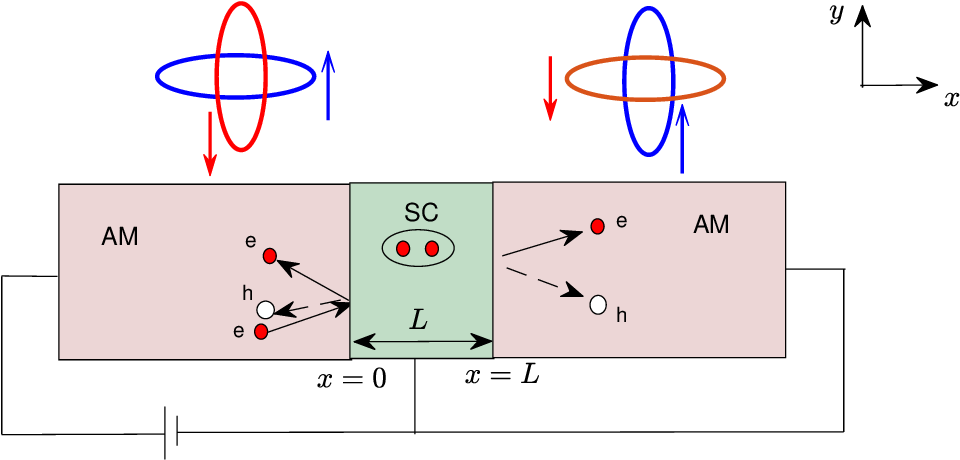}
 \caption{Schematic of the setup. An s-wave SC is sandwiched between two AMs that are rotated by $90^{\circ}$ with respect to each other. Bias is applied from the left AM, maintaining the SC and the right AM grounded. An electron incident from the left AM onto SC can reflect either as an electron or as a hole, or transmit through SC and emerge onto the right AM as an electron or as a hole. }\label{fig:schem}
\end{figure}  

 \section{Set-up}
 The Hamiltonian for an AM is given by
 \bea H_{\vec k} &=& -2t_0(\cos{k_xa}+\cos{k_ya})\si_0  \nn \\ &&+2t_J(\cos{k_xa}-\cos{k_ya})\si_z -\mu, \label{eq:ham0}\eea
 where $t_J$ is the spin and direction dependent hopping which characterizes the altermagnetic phase, $t_0$ is the hopping, $a$ is the lattice spacing and $\si_0,~\si_z$ are identity- and Pauli spin matrices. While most AMs are insulating, in this work by AM, we mean altermagnetic metals. In an earlier work~\cite{das2023}, we classified AMs  into  strong- or weak- phase depending upon whether $t_J>t_0$ or $t_J<t_0$. We stick to this convention for convenience. RuO$_2$ is an example for AM in the weak phase~\cite{smejkal22c} while Mn$_5$Si$_3$ is possibly an example for AM in the strong phase~\cite{mn5si3}.  We consider an AM-SC-AM junction arranged in such a way that the AM on the right is rotated by $90^{\circ}$ with respect to the AM on the left, as shown in Fig.~\ref{fig:schem}. We shall see that this helps in enhancing CAR. A bias is applied from the left AM, keeping the SC and the right AM grounded. We calculate the local conductivity $G_{LL}=dI_L/dV$ and the nonlocal conductivity $G_{RL}=dI_R/dV$, where $V$ is the bias, $I_{L}$ ($I_R$) is the current density in the left (right) AM, by Landauer-B\"uttiker approach~\cite{landauer1957r,buttiker1985m,soori17}. \\~
 
\section{Altermagnets in the strong phase}
 In this section, we consider the AMs to be in the strong phase by choosing $t_0=0$, $t_J>0$. This choice of parameters captures the essential physics of the setup in the strong phase of AMs and is not motivated by the experimental values of the parameters. The Hamiltonian for the AM on the left is obtained by expanding the Hamiltonian in eq.~\ref{eq:ham0} around the band bottom. The band bottoms for the two spins are located at different points in the Brillouin zone. For the AM on the left, the band bottom for $\ua$ [$\da$] -spin is at $(\pm \pi/a,0)$ [$(0,\pm\pi/a)$]. The Hamiltonian in the superconducting region mixes $\ua$ ($\da$)-spin electron with $\da$ ($\ua$)-spin hole. $\si_z$ commutes with the full Hamiltonian. Hence, we can work in the two sectors: (i) $\ua$-spin electron-$\da$-spin hole [$\ua_e$, $\da_h$], and (ii) $\da$-spin electron-$\ua$-spin hole [$\da_e$, $\ua_h$] separately. 
  
\subsection{($\ua_e$, $\da_h$) sector}
In the sector (i), the Hamiltonian can be written as $\sum_{\vec k}\Psi_{\vec k}^{\dagger}H_{\vec k}\Psi_{\vec k}$, where $\Psi_{\vec k}=[c_{\ua,k}, c^{\dagger}_{\da,-k}]^T$  and     
\begin{widetext}
\bea 
H_{\vec k} & =& \begin{cases} [t_J((k_xa \pm \pi)^2+k_y^2a^2)-\mu] \frac{\tau_z+\tau_0}{2} 
+ [t_J(k_x^2a^2+(k_ya \pm \pi)^2)-\mu]\frac{\tau_z-\tau_0}{2},~~{\rm for}~~ x<0,  \\ ~~ \\
  \Big[\frac{\hbar^2~(k_x^2+k_y^2)}{2~m}-\mu_s\Big]\tau_z + \Delta~\tau_x, ~~~{\rm for}~~ 0<x<L,    \\ ~~\\
  [t_J((k_xa \pm \pi)^2+k_y^2a^2)-\mu] \frac{\tau_z-\tau_0}{2} 
 + [t_J(k_x^2a^2+(k_ya \pm \pi)^2)-\mu] \frac{\tau_z+\tau_0}{2},~~{\rm for}~~x>L,
 \end{cases}
\label{eq:ham-up}
\eea
\end{widetext}
where $\tau_j$, for $j=0,x,z$ are the Pauli spin matrices acting on the particle-hole sector,  $\Delta$ is the superconducting gap, and $c_{\vec k,\si}$ annihilates an electron with wave vector $\vec k$ and spin $\si$. In eq.~\eqref{eq:ham-up}, the AM Hamiltonian is expanded around $k_{x}=\pm \pi/a$ and $k_y=\pm\pi/a$. When the range of $k_x$ ($k_y$) is taken to be $[-\pi/a,\pi/a]$, the Hamiltonian is expanded about $k_x=\pi/a$ ($k_y=\pi/a$) for positive $k_x$ ($k_y$) and about $k_x=-\pi/a$ ($k_y=-\pi/a$) for negative $k_x$ ($k_y$). 

 \begin{figure}[htb]
 \includegraphics[width=4.0cm,height=3.5cm]{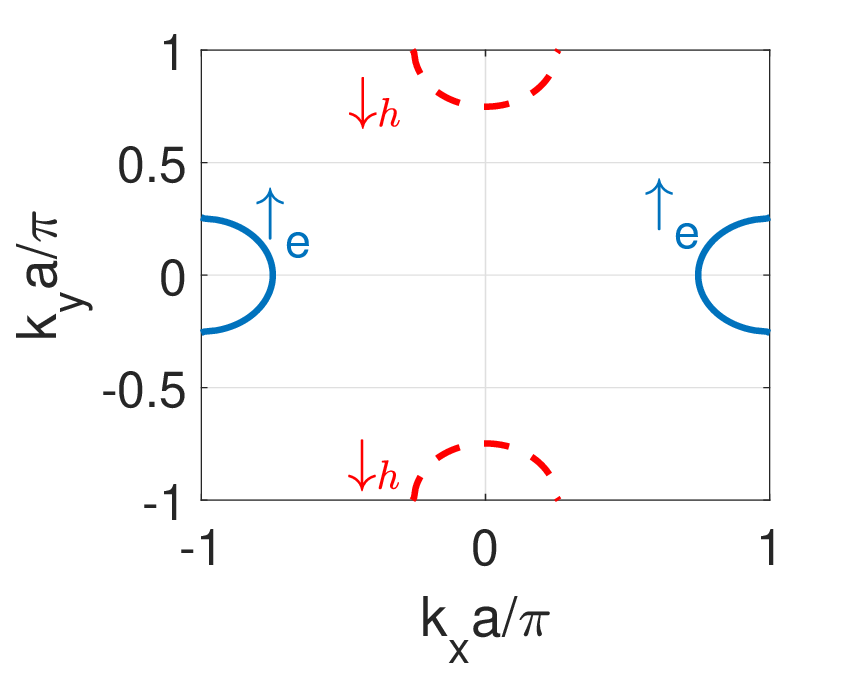}
 \includegraphics[width=4.0cm,height=3.5cm]{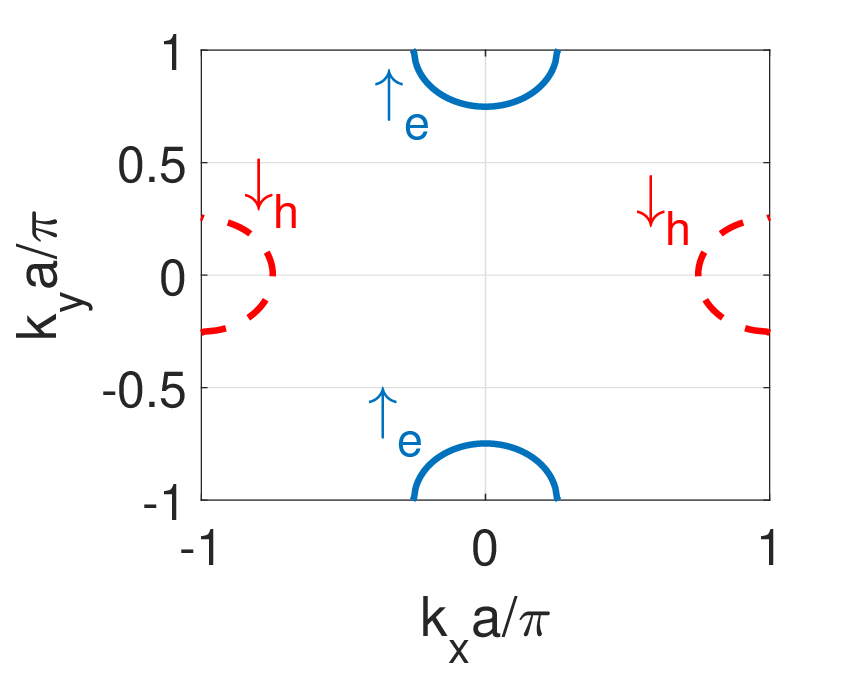}
  \caption{Schematic of the Fermi surfaces of  AMs on the left and the right in the strong phase. Blue solid line shows  $\ua_e$ and red dashed line shows  $\da_h$. }\label{fig:fs}
\end{figure}

Four processes can occur when an  electron is incident on the AM-SC-AM junction from the left AM. The electron with same spin can get reflected [electron reflection (ER)] or a hole  with opposite spin can get reflected [Andreev reflection (AR)]~\cite{btk} or an electron with same spin can transmit to the other AM [electron tunneling (ET)] or a hole with opposite spin can transmit to the other AM [crossed Andreev reflection (CAR)]~\cite{cht,beckmann04}.  Conservation of probability current density along $x$-direction  results in a competition among the four processes.  Suppression of any one of them is balanced by enhancement in  other processes.
 Since the system is translationally invariant along $y$-direction, matching or mismatching of $k_y$ in the dispersion relations on different sides determines which of the above phenomena is more likely to occur. For instance, we can see that in Fig.~\ref{fig:fs}, for $\ua$-spin electron incident from the left AM with a particular $k_y$, there exists $\da$-spin hole on the right AM with same $k_y$, but there exists no $\ua$-spin electron states with the same $k_y$ on the right AM. But $k_y$ has to be the same for all the states in the two regions. This means that while the $k_x$ corresponding to the $\da$-spin hole is real, $k_x$ corresponding to the $\ua$-spin electron is complex on the right AM. Since the states with real $k_x$ carry current and the states with complex $k_x$ carry no current, current on the right AM is carried by $\da$-spin holes alone.

 If on the right side, the crystallographic orientation of the AM is same as that on the left side, then for $k_y$ of incident up-spin electrons, and that of down-spin hole on the right AM do not match, which means CAR is not possible.  But $k_y$ values for the up-spin electrons on the left and the right AM regions match, resulting ET. Therefore, we notice that if we take two strong AMs whose crystallographic orientations are the same on the two sides of a SC, CAR does not occur. When the two AMs are oriented at $90^{\circ}$ to each other, CAR happens, but not ET. 
 
The eigenstates in AM regions are either purely electron-like or purely hole-like. The energy eigenvalues in the left AM for $\ua_e$ and $\da_h$ are $E_{e,\ua}=t_J[(k_{ex}a-\pi)^2+k_{ey}^2a^2]-\mu$ and $E_{h,\da}=-t_J[k_{hx}^2a^2+(k_{hy}a\pm\pi)^2]+\mu$ respectively. The eigenvectors for $\ua_e$ and $\da_h$ are $[1, 0]^T$ and $[ 0, 1]^T$ respectively. The eigenenergies in the right AM are $E_{e,\ua}=t_J[k_{ex}^2a^2+(k_{ey}a\pm \pi)^2]-\mu$ and  $E_{h,\da}=-t_J[(k_{hx}a-\pi)^2+k_{hy}^2a^2]+\mu$. Along $k_x$, we expand the Hamiltonian around $k_x=\pi/a$ and take the range of $k_x$ to be $[0,2\pi]$ for convenience. This does not change results. On the other hand, we take the range of $k_y$ to be $[-\pi/a,\pi/a]$. 

In the SC region, the dispersion is  $E=\pm \sqrt{[{\hbar^2}(q_x^2+q_y^2)/2m-\mu_s]^2+\Delta^2}$. The eigenstates are Bogoliubov–de Gennes (BdG) quasiparticles,  which have both the components: electron and hole.
In the SC region, when the bias is within the superconducting energy gap,  the BdG  states are evanescent modes and are equally  electron-like and hole-like. The eigenspinors are $[u_j, \Delta]^T$, where
\bea
u_j&=&E+\frac{\hbar^2}{2m}(q_{xj}^2+q_{y}^2)-\mu,~~{\rm where} ~~j=1,2,3,4,  \nn \\
q_{x1}&=&\sqrt{\frac{2m}{\hbar^2}[\mu + \sqrt{E^2-\Delta^2}]-q_y^2}, \nonumber\\
q_{x2}&=&-\sqrt{\frac{2m}{\hbar^2}[\mu + \sqrt{E^2-\Delta^2}]-q_y^2}, \nonumber \\
q_{x3}&=&+\sqrt{\frac{2m}{\hbar^2}[\mu - \sqrt{E^2-\Delta^2}]-q_y^2}, \nonumber \\
q_{x4}&=&-\sqrt{\frac{2m}{\hbar^2}[\mu - \sqrt{E^2-\Delta^2}]-q_y^2}. \label{eq:ujqj}
\eea

An electron incident on the AM-SC interface from the left, results in four processes which can happen as said above. The wave function corresponding to this in different regions has the form $\psi~e^{ik_{ey}y}$, where $\psi$ is given by 
\bea 
    \psi &=& e^{ik_{exr}x} \begin{bmatrix}
1 \\ 0 \\ \end{bmatrix}+r_{\ua_e}e^{i~k_{exl}x} \begin{bmatrix} 1 \\ 0 \\ 
\end{bmatrix}\nn \\&& +r_{\da_h}e^{i~k_{hx}x} \begin{bmatrix}  0 \\ 1\\ \end{bmatrix},~~~{\rm for}~~x<0 \nonumber \\
 &=& \sum_{j=1}^4B_je^{iq_{xj}x} \begin{bmatrix}
u_j \\  \Delta\\ \end{bmatrix},~~~~~~~{\rm for}~~0<x<L \nonumber \\ 
 &=& t_{\ua_e}e^{ik'_{ex}x} \begin{bmatrix} 1 \\ 0 \\ 
\end{bmatrix}+t_{\da_h}e^{-ik'_{hx\da}x} \begin{bmatrix}  0 \\ 1\\ \end{bmatrix},~~{\rm for}~~x>L
\label{eq:wf1}
\eea 
where $r_{\ua_e}, r_{\da_h},t_{\ua_e}$ and $t_{\da_h}$ are the coefficients for the processes: ER, AR, ET and CAR respectively. Here, $k_{exr}=\pi/a+k_e\cos{\theta}$,  and $k_{exl}=\pi/a-k_e\cos{\theta}$ denote the wave vector of  right moving electron and left moving electron respectively,  $k_{hx}=\sqrt{k_h^2-(k_{ey}-\pi~{\rm sign}(\theta)/a)^2}$ is the wave vector of hole associated with Andreev reflected hole,  $k'_{ex}=\sqrt{k_e^2-(k_{ey}-\pi~{\rm sign}(\theta)/a)^2}$,  $k'_{hx\da}=\pi/a-\sqrt{(k_{h}^2-k_{ey}^2)}$ are the wave vectors of electron and hole which are associated with ET and CAR respectively, where 
$k_ea=\sqrt{{(\mu+E)}/{t_J}}$ and $k_ha=\sqrt{{(\mu-E)}/{t_J}}$. 
Due to the translational invariance along y direction, we have $q_{y}=k_{ey}=k_e\sin{\theta}$. Here, we choose the parameters  so that $k_{hx}$ and $k'_{ex}$ are imaginary. This means that AR and ET are absent. $k_{hx}$ ($k'_{ex}$) has a negative (positive) imaginary part.

Now, to determine the scattering coefficients, boundary conditions are needed. These can be determined by demanding conservation of probability current density along $x$-direction. 
The  probability current densities along $x$ direction on the left AM- $J^P_{L,AM}$,  on the SC-  $J^P_{SC}$, and on the right AM- $J^P_{R,AM}$ are given by 

\bea 
J^P_{L,AM} &=& t_Ja^2[2{\rm Im}(\psi^{\dag}\tau_z\Do_x\psi)-\pi \psi^{\dag}(\tau_z+\tau_0)\psi/a]/\hbar, \nn \\   J^P_{SC}&=&~\hbar~{\rm Im}[\psi^{\dag}\tau_z\Do_x\psi]/m, \nn \\ J^P_{R,AM}&=&t_Ja^2[2{\rm Im}(\psi^{\dag}\tau_z\Do_x\psi)-\pi \psi^{\dag}(\tau_z-\tau_0)\psi/a]/\hbar \nn \\
&& \label{eq:str-prob-curr}
\eea 
The boundary conditions that conserve probability current density along $x$-direction are:
\bea 
\psi(0^-) &=& c\psi(0^+),~~ \nn \\   
\psi(L^-) &=& c\psi(L^+),~~ \nn \\ 
  \big(\f{\hbar^2}{2~m~a^2~t_J}\Do_x\psi-q_0\psi\big)_{0^+} &=& c\begin{pmatrix}\Do_x\psi_e-i\frac{\pi}{a}\psi_e\\\Do_x\psi_h \end{pmatrix}_{0^-}  ~~\nn \\  c\big(\f{\hbar^2}{2~m~a^2~t_J}\Do_x\psi+q_0\psi\big)_{L^-} &=& \begin{pmatrix}\Do_x\psi_e\\\Do_x\psi_h-i\frac{\pi}{a}\psi_h \end{pmatrix}_{L^+}
 \label{eq:bcup}\eea
    where $\psi=[ \psi_e,~ \psi_h ]^T$. The parameter $q_0$ used in the boundary conditions quantifies the strength of the delta-function barrier at the interface~\cite{soori23scat}. The parameter $c$ is real and dimensionless. Physically it corresponds to the strength of hopping from AM to SC in an equivalent lattice model~\cite{das2023}.  We choose $c=1$. Using these boundary conditions on the wave function having the form in eq.~\eqref{eq:wf1}, the scattering coefficients can be determined. In subsection~\ref{sec:results}, we will demonstrate the probability current conservation. 
 
Charge density is given by $\rho_c=e\psi^{\dag}\tau_z\psi$. Charge density does not commute with the Hamiltonian in the SC region, though it commutes with the Hamiltonian's in the AMs.   By using the continuity equation, we find charge current density on the left and the right AMs. The expressions for the charge current densities on the left and right AMs upon substituting the form of the wave function,   are given by 
 \bea
 I_{L,\ua_e,\da_h} &=& \frac{2 e t_J}{\hbar} [k_e \cos{\theta}(1-|r_{\ua_e}|^2)
 \nn \\
 I_{R,\ua_e,\da_h} &=& \frac{2 e t_J}{\hbar} [
 (k'_{hx\da}-\pi/a)|t_{\da_h}|^2] \label{eq:I-up}
 \eea 
\\
\subsection{($\da_e,\ua_h$) sector}
 In this sector, the calculation of currents can be done in a way similar to that followed in the previous subsection. For completeness, below we mention the Hamiltonian, boundary conditions, the scattering eigenfunction and the charge current density. The Hamiltonian for this sector is given by $\sum_{\vec k}\Psi_{\vec k}^{\dagger}H_{\vec k}\Psi_{\vec k}$, where $\Psi_{\vec k}=[c_{\da,k}, -c^{\dagger}_{\ua,-k}]^T$ and 
 \begin{widetext}
\bea 
H_{\vec k} &=& 
\begin{cases}
[t_J(k_x^2a^2+(k_ya \pm \pi)^2)-\mu] \frac{\tau_z+\tau_0}{2}  +[t_J((k_xa \pm \pi)^2+k_y^2a^2)-\mu] \frac{\tau_z-\tau_0}{2} ,~~{\rm ~for}~~x<0, \\ ~~\\
 \Big[\frac{\hbar^2~(k_x^2+k_y^2)}{2~m}-\mu_s\Big]\tau_z +\Delta~\tau_x, ~~~~~~~~{\rm for }~~ 0<x<L,    \\ ~~\\
  [t_J((k_xa \pm \pi)^2+k_y^2a^2)-\mu] \frac{\tau_z+\tau_0}{2} 
+ [t_J(k_x^2a^2+(k_ya \pm \pi)^2)-\mu] \frac{\tau_z-\tau_0}{2}, ~~{\rm for}~~x>L .
\end{cases}
\label{eq:ham2}
\eea 
\end{widetext}
The wave function corresponding to an electron incident from the left AM onto the interface at energy $E$ and angle of incidence $\theta$ has the form  $\psi~e^{ik_{ey}y}$, where $\psi$ is 

\bea 
    \psi &=& e^{ik_{ex}x} \begin{bmatrix}
1 \\ 0 \\ \end{bmatrix}+r_{\da_e}e^{-ik_{ex}x} \begin{bmatrix} 1 \\ 0 \\ 
\end{bmatrix}+ r_{\ua_h}e^{ik_{hx}x} \begin{bmatrix}  0 \\ 1\\ \end{bmatrix},\nn \\ &&~~~~~~{\rm for}~~x<0 \nonumber \\
 &=& \sum_{j=1}^4B_je^{iq_{xj}x} \begin{bmatrix}
u_j \\  \Delta\\ \end{bmatrix},~~~~~~~{\rm~for}~~0<x<L \nonumber \\
 &=& t_{\da_e}e^{ik'_{ex}x} \begin{bmatrix} 1 \\ 0 \\ 
\end{bmatrix}+t_{\ua_h}e^{ik'_{hx\ua}x} \begin{bmatrix}  0 \\ 1\\ \end{bmatrix},~~~{\rm for}~~x>L
\eea
Here, $k_{ex}=k_e\cos{\theta}$  denotes the wave vector of  right moving electron. $k_{hx}=\pi/a+\sqrt{k_h^2-k_{ey}^2}$ denotes the wave vector of hole in the region $x<0$.  $k_{ey}=\pi {\rm sign}(\theta)/a-k_e\sin{\theta}$  whereas $k'_{ex}=\pi/a+\sqrt{k_e^2-k_{ey}^2}$,  $k'_{hx\ua}=-\sqrt{k_h^2-(k_{ey}-\pi~{\rm sign}(\theta)/a)^2}$ stand for the wave vector of electron and hole in the right AM ($k_ea=\sqrt{{(\mu+E)}/{t_J}}$ and $k_ha=\sqrt{{(\mu-E)}/{t_J}}$). ${\rm Im}(k_{hx})<0$ and ${\rm Im}(k'_{ex})>0$.
The boundary conditions are
 \bea 
 \psi(0^-) &=& ~\psi(0^+),~~ \nn \\    \begin{pmatrix}\Do_x\psi_e\\\Do_x\psi_h-i\frac{\pi}{a}\psi_h \end{pmatrix}_{0^-} &=&  \big(\f{\hbar^2}{2~m~a^2~t_J}\Do_x\psi-q_0\psi\big)_{0^+} ~~\nn \\  \psi(L^-) &=& ~\psi(L^+),~~ \nn \\  \big(\f{\hbar^2}{2~m~a^2~t_J}\Do_x\psi+q_0\psi\big)_{L^-} &=& \begin{pmatrix}\Do_x\psi_e-i\frac{\pi}{a}\psi_e\\\Do_x\psi_h \end{pmatrix}_{L^+},
 \eea
  where $\psi=[\psi_e, \psi_h]^T$

 Charge current densities on the left and right AM  are given by
 \bea
 I_{L,\da e,\ua h} &=& \frac{2~e~t_J}{\hbar} [k_{ex}~(1-|r_{\da_e}|^2)
 ]
 \nonumber \\
 I_{R,\da e,\ua h} &=& \frac{2~e~t_J}{\hbar} [
 k'_{hx\ua}|t_{\ua_h}|^2] \nn \\ \label{I-dn}
 \eea

 \subsection{Conductivity}
 The total current densities on the two AMs are $I_{L}=I_{L,\ua e,\da h}+I_{L,\da e,\ua h}$ and $I_R= I_{R,\ua e,\da h}+ I_{R,\da e,\ua h}$. 
The differential conductivities on the two sides under a bias $V$ from left AM are given by 
 \bea
 G_{LL} &=& \frac{e^2}{2\pi h}~k_{e}\Bigg[\int_{-\pi/2}^{\pi/2}  (1-|r_{\ua_e}|^2) \cos\theta\,d\theta \nn \\&& + \int_{-\pi/2}^{\pi/2}(1-|r_{\da_e}|^2)~\cos\theta\,d\theta\Bigg] , \nn \\
  G_{RL} &=& \frac{e^2}{2\pi h}\Bigg[ \int_{-\theta_{c}}^{\theta_{c}}  (k'_{hx\da}-\pi/a)|t_{\da_h}|^2  d\theta + \nn \\&& \int_{-\theta_{c}}^{\theta_{c}}k'_{hx\ua}|t_{\ua_h}|^2 d\theta \Bigg], 
 \eea
 where $\theta_c=\sin^{-1}[{\rm min}(k_h/k_e,1)]$. 
 Here, $G_{LL}$ is called the local conductivity and $G_{RL}$ the nonlocal conductivity.  The local conductivity corresponds to the fraction of incident electrons from the left AM that get passed on to the SC and the right AM. Note that in this case, AR is completely suppressed. The nonlocal conductivity purely gets contribution from CAR (since ET does not contribute in this case) and is a measure of how good the CAR is. 

 \begin{figure}[htb]
 \includegraphics[width=8.0cm]{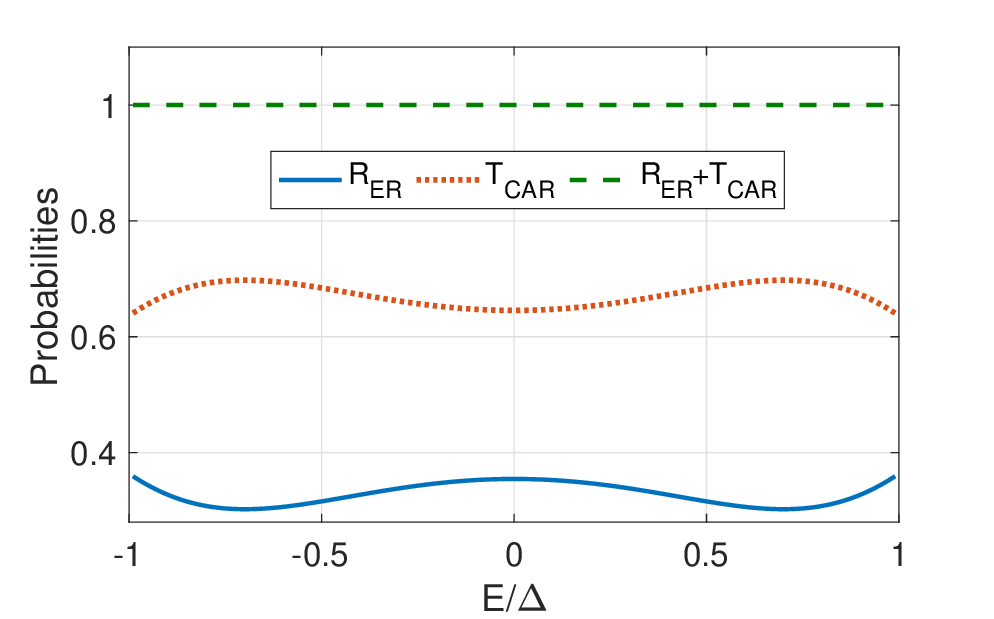}
 \caption{Probabilities of ER (solid line) and CAR (dotted line), followed by their sum (dashed line) versus energy for $\ua$-spin electron incident from the left AM with $\th=0^{+}$, $q_0=0$, $L=10.28a$, $m=\hbar^2/(a^2t_J)$, $\mu=0.2t_J$, $\Delta=0.1t_J$ and $\mu_s=2t_J$ are shown. For this case, the probabilities of AR and ET are zero.  The plot exhibits probability current conservation. } \label{fig:cons-str}
\end{figure}  

 \subsection{Results and Analysis}\label{sec:results}
To begin with, we shall examine the probability current conservation. Let us consider an up-spin electron incident from the left AM. By substituting the wave function [eq.~\eqref{eq:wf1}] into the expressions for the probability currents in eq.~\eqref{eq:str-prob-curr}, the condition  $J^P_{L,AM}=J^P_{R,AM}$ implies that $1=R_{ER}+T_{CAR}$, where $R_{ER}=|r_{\ua e}|^2$ and $T_{CAR}=|t_{\da h}|^2(\pi/a-k'_{hx\da})/k_e\cos{\th}$. In Fig.~\ref{fig:cons-str}, we plot the probabilities for ER and CAR: $R_{ER}$ and $T_{CAR}$, followed by their sum versus energy $E$, choosing the parameters: $q_0=0$, $L=10.28~a$, $m=\hbar^2/(a^2t_J)$, $\mu=0.2t_J$, $\Delta=0.1t_J$ and $\mu_s=2t_J$,  $L=10.28a$, and $\th\to 0^+$. 
 
 \begin{figure}[htb]
 \includegraphics[width=8.0cm]{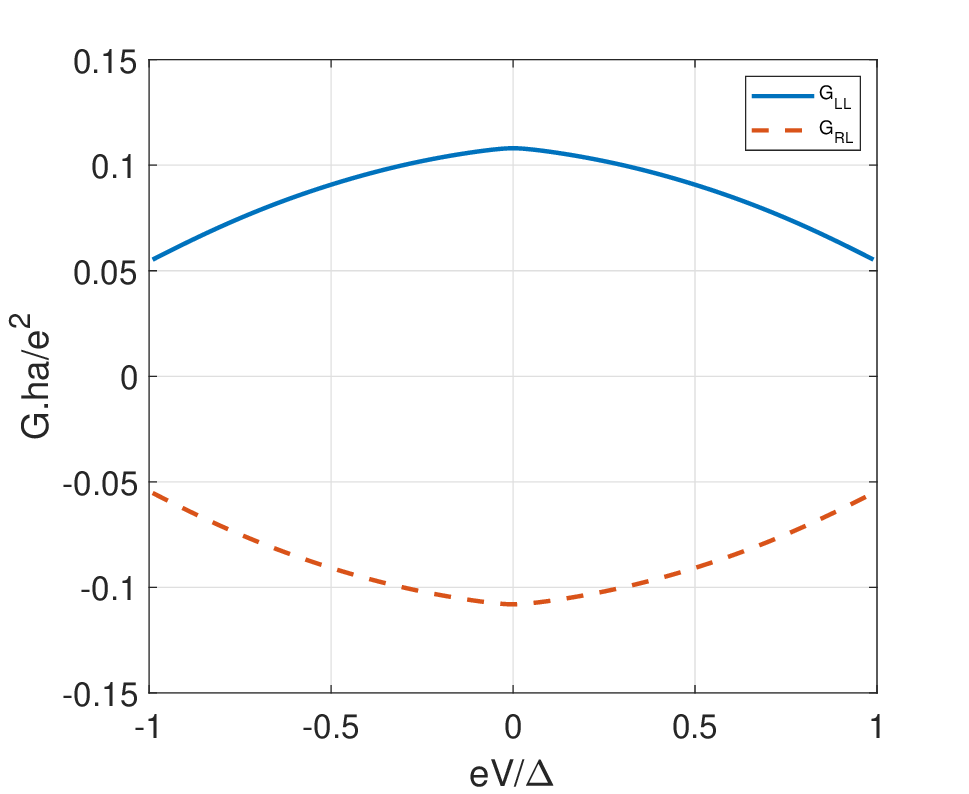}
 \caption{Local conductivity (blue line) and non-local conductivity (red dashed line) versus bias for $q_0=0$, $L=10.28a$, $m=\hbar^2/(a^2t_J)$, $\mu=0.2t_J$, $\Delta=0.1t_J$ and $\mu_s=2t_J$ are shown.} \label{fig:G1}
\end{figure}  

\begin{figure}[htb]
 \includegraphics[width=8.0cm]{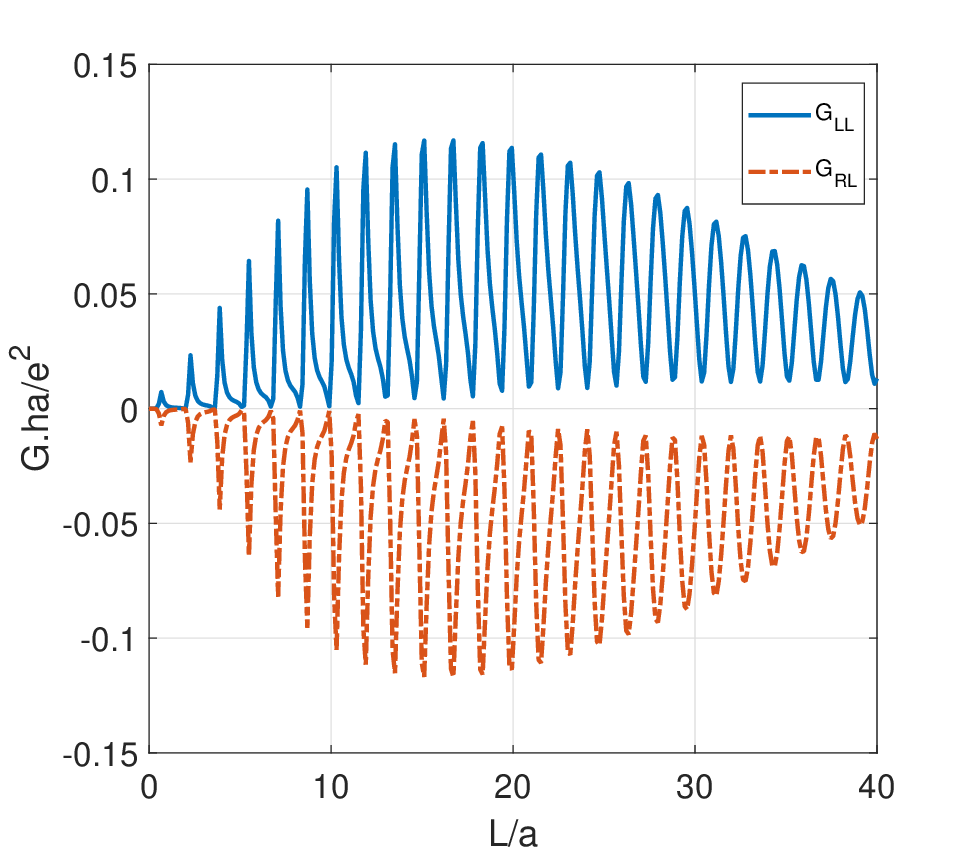}
 \caption{Local (solid blue line) and non-local (dash-dotted red line) conductivity versus the length of SC region are shown  for $q_0=0$, $eV=0$, $m=\hbar^2/(a^2t_J)$, $\mu=0.2t_J$, $\Delta=0.1t_J$ and $\mu_s=2t_J$. }\label{fig:G2}
\end{figure}

The local and nonlocal conductivities are numerically calculated following the procedure sketched in the previous subsections and plotted versus bias in Fig.~\ref{fig:G1} for the same set of parameters as earlier, except that $\th$ is no more a parameter now. $L$ is chosen to be $10.28a$ since CAR is enhanced for this choice of the length as can be seen in Fig.~\ref{fig:G2}. 
 We find that the local and non-local conductivity are exactly equal in magnitude and opposite in sign. This is because, there are only two processes that occur - CAR and ER. And from probability current conservation, the probability currents are equal on the left and the right AM. But, the charge current due to ER is $e$ times the probability current, whereas the charge current due to CAR is $-e$ times the probability current. This makes the charge currents on the two sides equal in magnitude  and opposite in sign. 
 The non-local conductivity shows a negative peak at zero energy. This is because for $E=0$, $k_e$ becomes exactly equal to $k_h$ i.e., $k_{ey}$ for all the incident electrons having momentum $k_e$,  matches  with $k_{hy}$ for all  the holes in the other AM having momentum $k_h$. So the conversion of electron to hole is maximum, giving rise to maximum CAR. But when energy $E\neq0$, there is a  mismatch in the transverse momentum values for $\ua_e$ and $\da_h$ resulting in lower conductivity. The non-local conductivity here is predominantly due to the  incident up-spin electrons. The contribution to conductivities from the incident down spin electrons is $10^{-9}$ times the contribution from incident up spin electrons.  The reason behind this is that for down-spin electron incidence $k_{ey}$ value is very large (near $\pi$) in comparison with $k_{ey}$ value for up-spin electron incidence (near $k_{ey}=0$). This results in a much smaller decay length for the evanescent modes in the SC region for the incident down spin electrons  compared to that for the incident up spin electrons. 
 
Now,  keeping all the parameters the same and fixing the bias to be at zero, we plot the conductivities versus length of the SC region in Fig.~\ref{fig:G2}. The conductivities show oscillations  due to Fabry-P\'erot interference~\cite{soori12,liang2001,soori19,sahu23} in the superconducting region. Within the SC gap, the wave numbers in the SC region are not purely real. The real part of the wave numbers is responsible for the Fabry-P\'erot interference. The Fabry-P\'erot interference condition is $\De L=\pi/k$, where $\De L$ is the separation between the consecutive peaks and $k$ is the real part of the wave number of the interfering mode in the SC. Here, we take the $k$ for the normal incidence, since the dominant contribution to the nonlocal transport is due to the electrons incident normal to the interface.  The value of $\De L$ calculated from this condition is 1.5695a in comparison to 1.6053a that is observed in the results in Fig.~\ref{fig:G2}. 
Further,  the peak value of the magnitude of non-local conductivity first increases with length, reaches a maximum value up to the order of $10^{-1}$ for superconducting length nearly equal to $15a$ and then gradually decreases. The global peak in the magnitude of nonlocal conductivity is due to maximum CAR which happens when the length of the SC is approximately inverse of the imaginary part of the wave number in the SC region, which is $20a$. This can be understood in the following way. Electron to hole conversion in SC is large within the SC gap. For small lengths of the SC, the electron to hole conversion is small. At small lengths, the electron to hole conversion probability increases with the length of the SC. But the wave function in the SC decays exponentially with the increase in the length of the SC and hence, for very large lengths of the SC, the converted hole does not reach the other AM. When the length of the SC is inverse of the imaginary part of the wave number, the electron to hole conversion is large and the wave function in the SC is optimum at the right SC-AM interface for the hole to exit into the right AM.  The nonlocal conductivity is always  negative for AMs in the strong phase chosen with appropriate crystallographic orientation, signalling dominant CAR  with zero contribution from ET. We numerically find that the contribution to the current from ET is zero, which can be understood by looking at Fig.~\ref{fig:fs}. Due to mismatch of $k_y$ for the electron states on the two AMs, $k_x$ on the right AM for $\ua$-spin electron is complex and hence the $\ua$-spin electron does not carry any current in the right AM.   This is in contrast to other schemes of enhancement of CAR, wherein the contribution to nonlocal conductivity from ET is typically nonzero~\cite{soori17,nehra19,beckmann04,russo2005experimental}. It may be noted here that there exist proposals in literature, where 100$\%$ CAR can be achieved in principle~\cite{Zhang19,Jakobsen21,Zhao}. 

\begin{figure}[htb]
 \includegraphics[width=4.0cm,height=3.8cm]{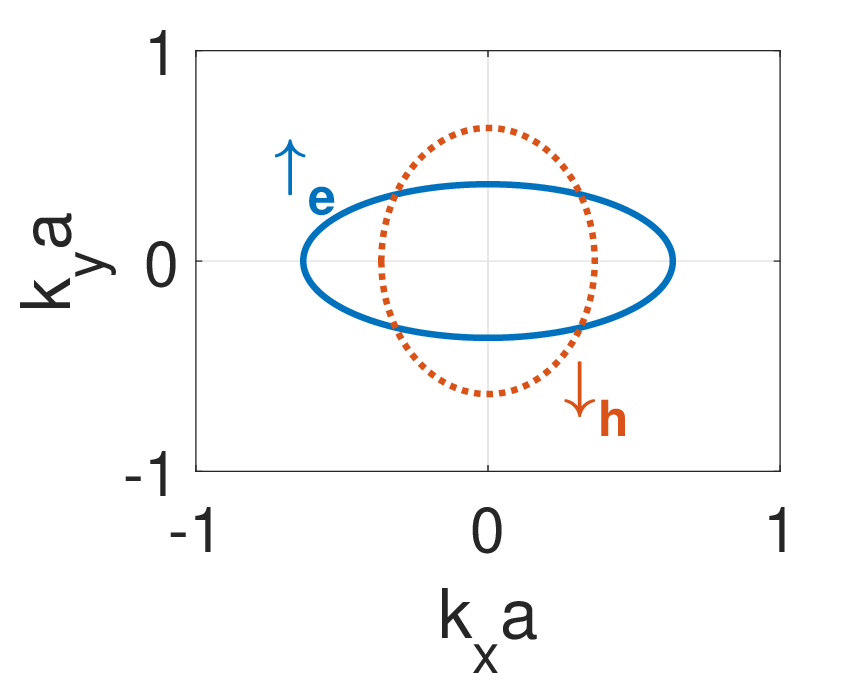}
 \includegraphics[width=4.0cm,height=3.8cm]{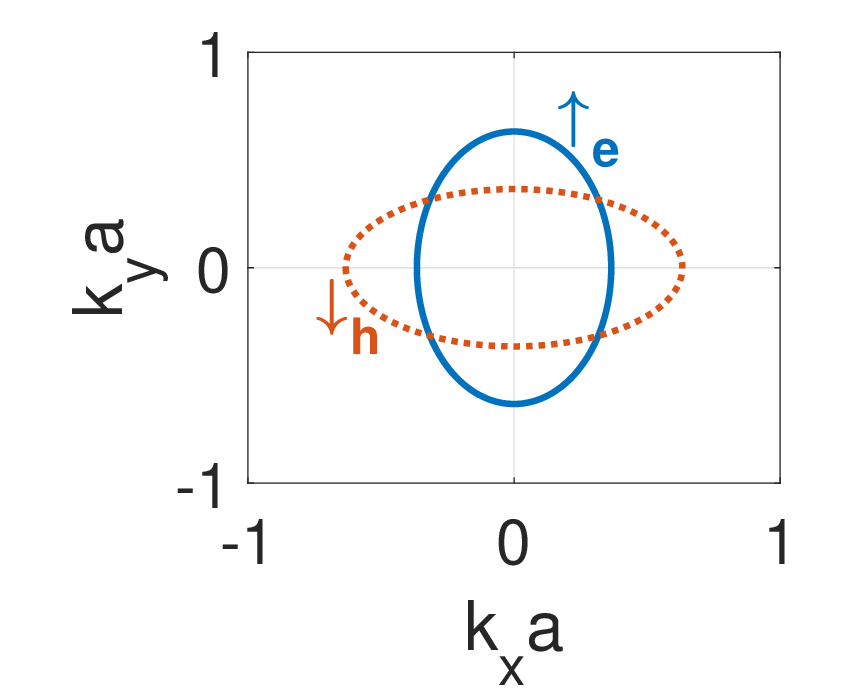}
 \caption{Fermi surfaces of  AMs on the left and the right in weak phase, blue solid line showing  $\ua_e$ and red dotted line shows  $\da_h$ for $t_J=0.5t_0$ and $\mu=0.2t_0$}\label{fig:FS3}
\end{figure}

 \section{Altermagnets in weak phase}
In this section, we choose the AMs to be in the weak phase by choosing  $t_0>t_J\ge 0$. Unlike in the strong phase,  the band bottoms for both the spins  are located at $k_x=k_y=0$. But the Fermi surfaces do not overlap, rather they intersect each other due to their anisotropic behavior. As a result, for a given spin, all the $k_y$ values for the electrons of AMs on either sides of the SC do not match. However, the $k_y$ values for electrons of spin $\si$ on one side match with $k_y$ values for holes of spin $\bar{\si}$  to a much larger extent. This can be seen from Fig.~\ref{fig:FS3}. Therefore, CAR is favored over ET on an average.

\subsection{($\ua_e,\da_h$) sector}
In this sector, the Hamiltonian can be written similar to ($\ua_e$, $\da_h$) case in strong phase. The difference is that now the dispersion is expanded around $k_x=0,k_y=0$ as the band bottom lies there.  The Hamiltonian is: 
\begin{widetext}
    
\bea
H_{\vec k} &=&
\begin{cases}
[(t_0-t_J)~k_x^2a^2+(t_0+t_J)~k_y^2a^2-\mu]\frac{\tau_z+\tau_0}{2} 
+ [(t_0+t_J)~k_x^2a^2+(t_0-t_J)~k_y^2a^2)-\mu]\frac{\tau_z-\tau_0}{2}, ~~~{\rm for}~~x<0 \\ ~~\\
 \Big[\frac{\hbar^2~(k_x^2+k_y^2)}{2~m}-\mu_s\Big]\tau_z + \Delta~\tau_x, ~~~~{\rm for}~~~ 0<x<L    \\~~\\
  [(t_0+t_J)~k_x^2a^2+(t_0-t_J)~k_y^2a^2-\mu]\frac{\tau_z+\tau_0}{2}  
+ [(t_0-t_J)~k_x^2a^2+(t_0+t_J)~k_y^2a^2)-\mu]\frac{\tau_z-\tau_0}{2},~~~{\rm for}~~x>L
\end{cases}
\label{eq:ham-weak-up}
\eea

\end{widetext}
In Fig.~\ref{fig:FS3},  the Fermi surfaces for  AMs on the two sides are shown. 
The wave function corresponding to the Hamiltonian in different regions possesses the form $\psi~e^{ik_{ey}y}$, where $\psi$ is given by
\bea
    \psi
    &=& e^{ik_{ex\ua}x} \begin{bmatrix}
1 \\ 0 \\ \end{bmatrix}+r_{\ua_e}e^{-ik_{ex\ua}x} \begin{bmatrix} 1 \\ 0 \\
\end{bmatrix}  +r_{\da_h}e^{ik_{hx\da}x} \begin{bmatrix}  0 \\ 1\\ \end{bmatrix}, \nn \\ &&~~~~{\rm ~ for ~} x<0, \nonumber \\ 
 &=& \sum_{j=1}^4B_je^{iq_{xj}x} \begin{bmatrix}
u_j \\  \Delta\\ \end{bmatrix},~~~~~~~~{\rm for~~}0<x<L, \nonumber \\
 &=& t_{\ua_e}e^{ik'_{ex\ua}x} \begin{bmatrix} 1 \\ 0 \\
\end{bmatrix}+t_{\da_h}e^{ik'_{hx\da}x} \begin{bmatrix}  0 \\ 1\\ \end{bmatrix}, ~~~~~~{\rm for} ~~~~~x>L, \nn \\ && 
\eea
and $r_{\ua_e}, r_{\da_h},t_{\ua_e}$ and $t_{\da_h}$ are the scattering coefficients for ER, AR, ET and CAR respectively. Here, $k_{ex\ua}a=\sqrt{(E+\mu)/(t_0-t_J)}\cos{\theta}$ is the wave vector associated with right moving electron, $k_{ey}a=\sqrt{(E+\mu)/(t_0+t_J)}\sin{\theta}$ is the component of wave vector of electron along $y$-direction. 
$k_{hx\da}a=\sqrt{\big[(\mu-E)-(t_0-t_J)~k_{ey}^2a^2\big]/(t_0+t_J)}$ is the wave vector associated with reflected hole in left AM whereas 
$k'_{ex\ua}a=\sqrt{\big[(\mu+E)-(t_0-t_J)~k_{ey}^2a^2\big]/(t_0+t_J)}$ and  $k'_{hx\da}a=-\sqrt{\big[(\mu-E)-(t_0+t_J)~k_{ey}^2a^2\big]/(t_0-t_J)}$ stand for the wave vectors for the transmitted  electron and transmitted hole respectively. Whenever any of these wave numbers turn out to be complex, the square root is taken so that the wave decays to zero at $x\to\pm\infty$. 

The  probability current density along $x$ direction on: the left AM- $J^P_{L,AM}$,  central SC- $J^P_{SC}$ and the right AM- $J^P_{R,AM}$ are given by
\begin{widetext}
\bea
J^P_{L,AM}&=&\f{(t_0-t_J)a^2[{\rm Im}(\psi^{\dag}(\tau_z+\tau_0)\Do_x\psi)] - (t_0+t_J)a^2[{\rm Im}(\psi^{\dag}(\tau_z-\tau_0)\Do_x\psi)]}{\hbar} \nn \\ J^P_{SC}&=&\f{\hbar ~{\rm Im}[\psi^{\dag}\tau_z\Do_x\psi]}{m}\nn \\ 
J^P_{R,AM}&=&\f{(t_0+t_J)a^2[{\rm Im}(\psi^{\dag}(\tau_z+\tau_0)\Do_x\psi)] - (t_0-t_J)a^2[{\rm Im}(\psi^{\dag}(\tau_z-\tau_0)\Do_x\psi)]}{\hbar} \eea
\end{widetext}
From probability current conservation on the two sides, we can find the boundary conditions which will ultimately help us in determining the scattering coefficients. We choose the following  boundary conditions:
  \bea 
 \psi(0^-) &=& ~\psi(0^+),~~ \nn \\ \Do_x\psi|_{0^-} &=&\begin{pmatrix}
     \f{\hbar^2}{2ma^2(t_0-t_J)}\Do_x\psi_e-q_0\psi_e\\\f{\hbar^2}{2ma^2(t_0+t_J)}\Do_x\psi_h-q_0\psi_h
 \end{pmatrix}_{0^+}  ~~ \nn \\  \psi(L^+) &=& ~\psi(L^-),~~ \nn \\     \Do_x\psi|_{L^+} &=&  \begin{pmatrix}
     \f{\hbar^2}{2ma^2(t_0+t_J)}\Do_x\psi_e+q_0\psi_e\\\f{\hbar^2}{2ma^2(t_0-t_J)}\Do_x\psi_h+q_0\psi_h \end{pmatrix}_{L^-}
 \eea \\ 
Charge current density in the left and the right AM are given by the following expressions:
  \bea
 I_{L,\ua_e,\da_h} &=& \frac{2 e}{\hbar} [(t_0-t_J)k_e \cos{\theta}(1-|r_{\ua_e}|^2)\nn \\&& +(t_0+t_J){\rm Re}(k_{hx\da})|r_{\da_h}|^2] \nn \\
 I_{R,\ua_e,\da_h} &=& \frac{2 e}{\hbar} [(t_0+t_J){\rm Re}(k'_{ex\ua}) |t_{\ua_e}|^2 \nn \\&& +(t_0-t_J){\rm Re}(k'_{hx\da})|t_{\da_h}|^2] 
 \eea 
 
 \subsection{($\da_e,\ua_h$) sector}
 We use a method similar to that in the previous subsection to find the current density in this sector. So we start with the  Hamiltonian:
 \begin{widetext}
 \bea 
H_{\vec k} &=& \begin{cases}
     [(t_0+t_J)~k_x^2a^2+(t_0-t_J)~k_y^2a^2-\mu]\frac{\tau_z+\tau_0}{2}  
+ [(t_0-t_J)~k_x^2a^2+(t_0+t_J)~k_y^2a^2)-\mu]\frac{\tau_z-\tau_0}{2}, ~~~{\rm for}~~x<0  \\~\\
 \Big[\frac{\hbar^2~(k_x^2+k_y^2)}{2~m}-\mu_s\Big]\tau_z + \Delta~\tau_x, ~~~~~~~~~{\rm for}~~~ 0<x<L   \\~\\
   [(t_0-t_J)~k_x^2a^2+(t_0+t_J)~k_y^2a^2-\mu]\frac{\tau_z+\tau_0}{2}  
+ [(t_0+t_J)~k_x^2a^2+(t_0-t_J)~k_y^2a^2)-\mu]\frac{\tau_z-\tau_0}{2},~~~{\rm for}~~x>L
\end{cases}
\eea
 \end{widetext}
The wave function for the system has the form $\psi~e^{ik_{ey}y}$ where
\bea
    \psi &=& e^{ik_{ex\da}x} \begin{bmatrix}
 1 \\ 0 \\ \end{bmatrix}+r_{\da_e}e^{-ik_{ex\da}x} \begin{bmatrix}  1 \\ 0 \\ 
\end{bmatrix}  +r_{\ua_h}e^{ik_{hx\ua}x} \begin{bmatrix}  0 \\ 1 \\ \end{bmatrix}, \nn \\&& ~~~~~~~~~~~~~~~~~~~~~~~~~~~~~~~~~~~~~~~~~~~~{\rm for}~~ x<0 \nonumber \\
 &=& \sum_{j=1}^4B_je^{iq_{xj}x} \begin{bmatrix}
u_j \\  \Delta\\ \end{bmatrix},~~~~~~~~~~~{\rm for}~~0<x<L \nonumber \\
 &=& t_{\da_e}e^{ik'_{ex\da}x} \begin{bmatrix} 0 \\ 1 \\
\end{bmatrix}+t_{\ua_h}e^{ik'_{hx\ua}x} \begin{bmatrix}   1 \\ 0\\ \end{bmatrix},~~{\rm for}~~ x>L,
\eea 
and $k_{ex\da}a=\sqrt{(E+\mu)/(t_0+t_J)}~\cos{\theta}$,\\  $k_{ey}a=\sqrt{(E+\mu)/(t_0-t_J)}~\sin{\theta}$,  \\ 
$k_{hx\ua}a=\sqrt{\big[(\mu-E)-(t_0+t_J)~k_{ey}^2a^2\big]/{(t_0-t_J)}},\\
k'_{ex\da}a=\sqrt{[(E+\mu)-(t_0+t_J)~k_{ey}^2a^2]/{(t_0-t_J)}}$,\\ $k'_{hx\ua}a=-\sqrt{[(\mu-E)-(t_0-t_J)~k_{ey}^2a^2]/{(t_0+t_J)}}$. \\
All the above terms have the same meaning as mentioned in the previous subsection.
The boundary conditions for this sector are given by   
  \bea 
 \psi(0^-) &=& ~\psi(0^+),~~ \nn \\ \Do_x\psi|_{0^-} &=&\begin{pmatrix}
     \f{\hbar^2}{2ma^2(t_0+t_J)}\Do_x\psi_e-q_0\psi_e\\\f{\hbar^2}{2ma^2(t_0-t_J)}\Do_x\psi_h-q_0\psi_h
 \end{pmatrix}_{0^+}  ~~ \nn \\  \psi(L^+) &=& ~\psi(L^-),~~ \nn \\     \Do_x\psi|_{L^+} &=&  \begin{pmatrix}
     \f{\hbar^2}{2ma^2(t_0-t_J)}\Do_x\psi_e+q_0\psi_e\\\f{\hbar^2}{2ma^2(t_0+t_J)}\Do_x\psi_h+q_0\psi_h \end{pmatrix}_{L^-}
 \eea 

Charge current density in the left and right AM are given by the following expressions:
  \bea
 I_{L,\da_e,\ua_h} &=& \frac{2 e}{\hbar} [(t_0+t_J)k_e \cos{\theta}(1-|r_{\da_e}|^2)\nn \\&& +(t_0-t_J){\rm Re}(k_{hx\ua})|r_{\ua_h}|^2] \nn \\
 I_{R,\da_e,\ua_h} &=& \frac{2 e}{\hbar} [(t_0-t_J){\rm Re}(k'_{ex\da}) |t_{\da_e}|^2 \nn \\&& +(t_0+t_J){\rm Re}(k'_{hx\ua})|t_{\ua_h}|^2] 
 \eea \\

\subsection{Conductivity}
The total current densities on the two AMs are $I_{L}=I_{L,\ua e,\da h}+I_{L,\da e,\ua h}$ and $I_R= I_{R,\ua e,\da h}+ I_{R,\da e,\ua h}$. Unlike in the case of AMs in the strong phase, here the local conductivity draws contributions from ER and AR. Similarly, the nonlocal conductivity gets contributions from ET and CAR. Negative value of nonlocal conductivity means that CAR overpowers ET.
The differential conductivities on the two sides under a bias $V$ from left AM are given by
\begin{widetext}

   \bea
 G_{LL} &=& 
 \frac{e^2}{2\pi h}\Bigg[\sqrt{\frac{t_0-t_J}{t_0+t_J}}\int_{-\pi/2}^{\pi/2}~k_{ex\ua} (1-|r_{\ua_e}|^2) d\theta +\sqrt{\frac{t_0+t_J}{t_0-t_J}}\int_{-\theta_{h\da }}^{\theta_{h\da }}k_{hx\da} (|r_{\da_h}|^2) d\theta \nn \\&& +\sqrt{\frac{t_0+t_J}{t_0-t_J}}\int_{-\pi/2}^{\pi/2}~k_{ex\da} (1-|r_{\da_e}|^2) d\theta +\sqrt{\frac{t_0-t_J}{t_0+t_J}}\int_{-\theta_{h\ua}}^{\theta_{h\ua}}k_{hx\ua} (|r_{\ua_h}|^2) d\theta\Bigg] ,  \\ 
&& {\rm where~~} 
\theta_{h\da }=\sin^{-1}\Bigg[{\rm min}\Big[\f{(\mu-E)(t_0+t_J)}{(\mu+E)(t_0-t_J)},1\Big]\Bigg], ~~ \theta_{h\ua}=\sin^{-1}\Bigg[{\rm min}\Big[\f{(\mu-E)(t_0-t_J)}{(\mu+E)(t_0+t_J)},1\Big]\Bigg],\nn \\
 G_{RL} &=& \frac{e^2}{2\pi h}\Bigg[\sqrt{\frac{t_0+t_J}{t_0-t_J}}\int_{-\pi/2}^{\pi/2}k'_{ex\ua} (|t_{\ua_e}|^2) d\theta  -\sqrt{\frac{t_0-t_J}{t_0+t_J}}\int_{-\theta_{h}}^{\theta_{h}}k'_{hx\da} (|t_{\da_h}|^2) d\theta \nn \\&& +\sqrt{\frac{t_0-t_J}{t_0+t_J}}\int_{-\th_e}^{\th_e}k'_{ex\da}(|t_{\da_e}|^2) d\theta  -\sqrt{\frac{t_0+t_J}{t_0-t_J}}\int_{-\theta_{h}}^{\theta_{h}}k'_{hx\ua}(|t_{\ua_h}|^2) d\theta\Bigg],  \\ \nn 
 && {\rm where~~} \th_e=\sin^{-1}[\sqrt{(t_0-t_J)/(t_0+t_J)}], ~~\theta_h=\sin^{-1}[{\rm min}\{(\mu-E)/(\mu+E),1\}]. 
 \eea
\end{widetext}

  \begin{figure}[htb]
 \includegraphics[width=8cm]{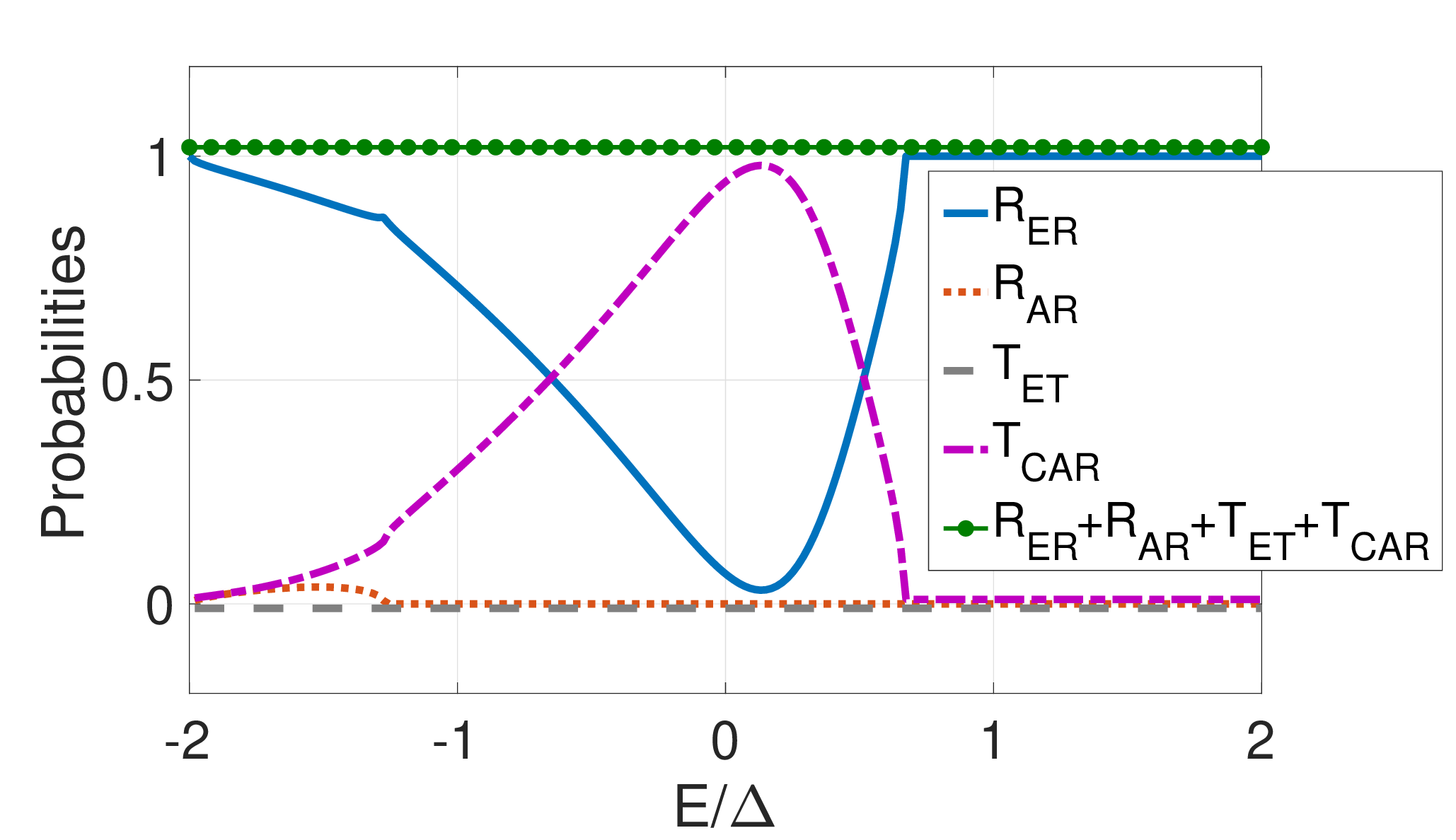}
\caption{Probabilities of ER (solid line), AR (dotted line), ET (dashed line), CAR (dot-dash line) and their sum (flat points line) versus energy $E$ for a $\da$-spin electron incident from the left AM with $\th=\pi/4$, $q_0=0$, $L=7a$,$t_J=0.8t_0$ $m=\hbar^2/(a^2t_0)$, $\mu=0.2t_0$, $\Delta=0.1t_0$ and $\mu_s=2t_0$ are shown.} \label{fig:cons-weak}
\end{figure}

 \subsection{Results and Analysis}
We begin this subsection by  demonstrating probability current conservation. For the weak phase, all four processes - ER, AR, ET and CAR contribute to probability current. Let us consider a down spin electron incident from the left AM. It can be shown that the probabilities of ER, AR, ET and CAR are respectively given by
\bea 
R_{ER} &=& |r_{\da e}|^2, \nn \\ 
R_{AR} &=&  \f{(t_0-t_J)|{\rm Re}(k_{h\ua}) |r_{\ua h}|^2}{(t_0+t_J)k_e\cos{\th}} ,\nn \\ 
T_{ET} &=&  \f{(t_0-t_J){\rm Re}(k'_{ex\da})|t_{e\da}|^2}{(t_0+t_J)k_e\cos{\th}} ,     \nn \\ 
T_{CAR} &=&  \f{|{\rm Re}(k'_{hx\ua})|~|t_{\ua h}|^2}{k_e\cos\th}  .
\eea
In Fig.~\ref{fig:cons-weak}, we plot these probabilities and their sum versus energy for  $\th=\pi/4$, $q_0=0$, $L=7a$,$t_J=0.8t_0$ $m=\hbar^2/(a^2t_0)$, $\mu=0.2t_0$, $\Delta=0.1t_0$ and $\mu_s=2t_0$. It can be seen that the probabilities for the four processes add up to $1$. Though for the angle of incidence $\th=\pi/4$, the down-spin electron exhibits almost perfect CAR near zero energy for this choice of parameters. 
 
  \begin{figure}[htb]
 \includegraphics[width=8cm]{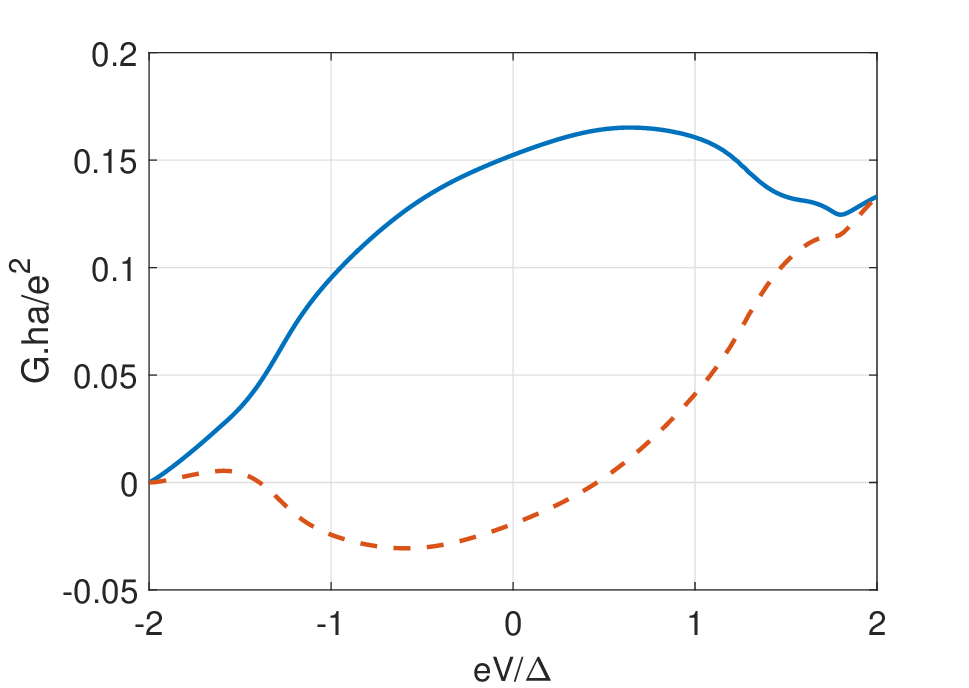}
\caption{Local conductivity (blue line) and non-local conductivity (red dashed line) versus bias for $q_0=0$, $L=7a$,$t_J=0.8t_0$ $m=\hbar^2/(a^2t_0)$, $\mu=0.2t_0$, $\Delta=0.1t_0$ and $\mu_s=2t_0$ are shown.} \label{fig:G3}
\end{figure}
In Fig.~\ref{fig:G3}, we plot the two conductivities versus bias and find that  CAR dominates over ET for a certain range of bias within the SC gap. Because, within the superconducting gap, the electron to hole conversion has higher probability.
 \begin{figure}[htb]
 \includegraphics[width=8.0cm]{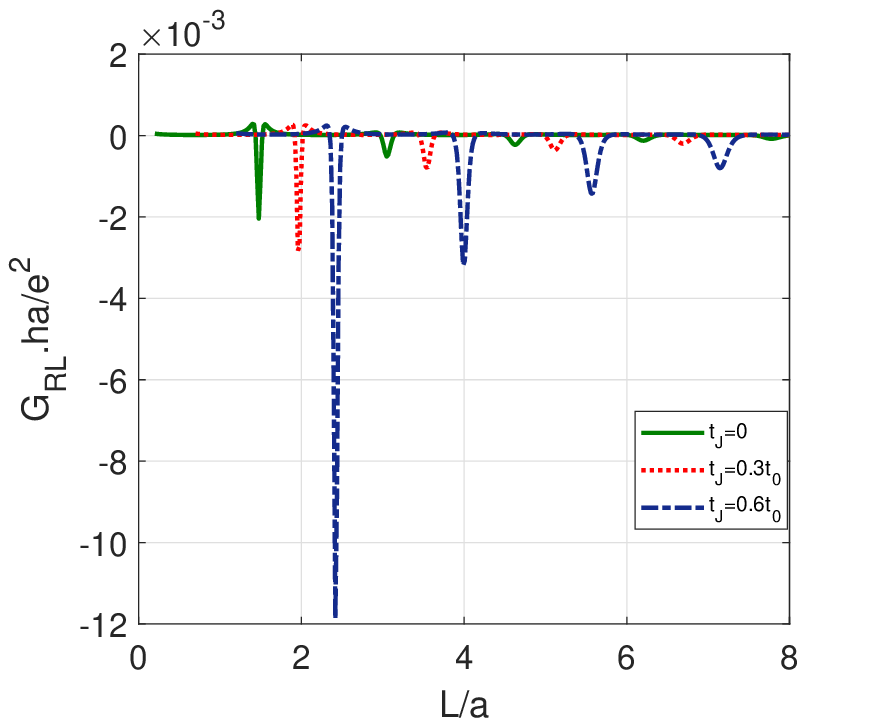}
 \caption{Non-local conductivity versus length of SC for $t_J=0$ (green solid line), $t_J=0.3t_0$ (red dotted-line), $t_J=0.6t_0$(blue dash-dot line), $E=0$, $q_0=10t_0$, $m=\hbar^2/(a^2t_0)$, $\mu=0.2t_0$, $\Delta=0.1t_0$, $\mu_s=2t_0$. Red dotted  and blue dash-dot curves are shifted by 0.5 and 1.0 units  along $x$-axis respectively for clarity.}\label{fig:G4}
\end{figure} 
In Fig.\ref{fig:G4}, the variation of non-local conductivity with  length for different values of $t_J$ is shown. $t_J=0$ corresponds to the absence of altermagnetic phase, and the leads behave as normal metal leads. We notice that when $t_J\neq0$  the non-local conductivity is higher in magnitude as compared to when $t_J=0$. Thus, we find that  AM helps to enhance CAR in comparison to a normal metal. The larger the value of $t_J$ within the range $0<t_J<t_0$, the larger is the non-local conductivity in magnitude. The non-local conductivity plot shows a negative peak for some definite values of length, and this peak is periodic in nature. The reason behind this is that the wave gets multiply reflected back-and-forth within the SC region and picks up a phase of  $2{\rm Re}(q_x)L$ in one round of back-and-forth reflection, where $q_x$ is the wave number in the SC region. According to the Fabry-P\'erot interference condition,  when this phase difference is integral multiple of $2\pi$, we get constructive interference resulting in peaks. The separation between two consecutive negative peaks can be calculated by $\Delta L={\pi}/{{\rm Re}(q_x)}$. The value of $\Delta L$ calculated from this formula is $1.57a$ and the value observed in Fig.~\ref{fig:G4} is $\sim 1.58a$.

 \section{Experimental relevance}
 Let us see  how the parameters $t_0$ and $t_J$ used in the model  are connected to real materials, taking the example of  RuO$_2$ - a well established AM. In RuO$_2$, the bandwidth and the spin splitting  are  of the same order of magnitude $\sim 1 eV$~\cite{papaj,smejkal22c}. However, the SC gap in conventional SCs like NbSe$_2$ is of the order of a few meV making the ratio of $\De/t_0\sim 10^{-3}$. The nonlocal conductivity is not negative for such a choice of parameters. To get negative nonlocal conductivity, $\De/t_0$ needs to be of the order of $0.1$ maintaining $t_J\sim t_0$. This can be achieved by choosing Mn$_5$Si$_3$ for AM which has a spin splitting of $150$~meV~~\cite{mn5si3}. For this material, the estimated values of $t_J\simeq 2t_0 \sim 150meV$. Some of the iron based SCs have a larger SC gap~\cite{Xu2012,Wang2011,Qian2011,Zhang2011,Mou2011}. In particular, the material (Tl$_{0.58}$Rb$_{0.42}$)Fe$_{1.72}$Se$_2$
has an SC gap of $15$meV~~\cite{Mou2011}. The choice of such iron based SC along with the AM Mn$_5$Si$_3$ will give us $\De/t_0\sim 0.1$ along with  $t_J\gtrsim t_0$ which is suited to test the predictions of our work.    
 \section{Conclusion}
We propose to employ the newly discovered AMs to enhance and detect crossed Andreev reflection across an s-wave SC. When the two AMs are rotated by $90^{\circ}$ with respect to each other, CAR can be enhanced  significantly.  We calculated the local and nonlocal conductivities for AM-SC-AM junctions.  We find that when the AMs are in the strong phase, the nonlocal transport can be completely dominated by CAR with zero contribution from ET. On the other hand, when the AMs are in the weak phase, both ET and CAR contribute to the nonlocal conductivity. But, for certain choice of parameters, we can get larger contribution from CAR than from ET. 
The nonlocal conductivity shows extrema for certain values of the length of the SC which is rooted in Fabry-P\'erot type interference. 
With a  careful choice of the length of the SC, the system can be tuned to enhance CAR.
Our results will be useful in development of superconducting devices based on the phenomenon of CAR.
 
\acknowledgements
We thank Dhavala Suri for illuminating discussions. SD and AS thank  SERB Core Research grant (CRG/2022/004311) for financial support. AS thanks the funding from University of Hyderabad Institute of Eminence PDF. 

\bibliography{ref_almag}
\end{document}